\documentclass[preprintnumbers,superscriptaddress,showkeys,showpacs,byrevtex]{revtex4}
\usepackage{amsmath,amsfonts,amssymb,amscd,amsxtra,amsthm}
\usepackage{graphicx}
\usepackage{bm}
\usepackage{amsmath}
\usepackage{multirow}

\newcommand {\be} {\begin{equation}}
\newcommand {\ba} {\begin{eqnarray}}
\newcommand {\ee} {\end{equation}}
\newcommand {\ea} {\end{eqnarray}}
\begin{document}
\preprint{CYCU-HEP-19-04}
\title{Kaon Multiplicities of Semi-inclusive DIS and the Fragmentation Functions }
%--------------------------------------------------
\author{Chung Wen Kao}
\email[E-mail: ]{cwkao@cycu.edu.tw}
\affiliation{Department of Physics and Center for High Energy Physics, Chung-Yuan Christian University, Chung-Li 32023, Taiwan}
%-----------------------------------------------------------
\author{Dong-Jing Yang}
\email[E-mail: ]{djyang@std.ntnu.edu.tw}
\affiliation{Department of Physics, National Taiwan Normal University,
Taipei 10610, Taiwan}
%-----------------------------------------------------
\author{Wen Chen Chang}
\email[E-mail: ]{changwc@phys.sinica.edu.tw}
\affiliation{Institute of Physics, Academia Sinica, Taipei 11529, Taiwan.}
\date{\today}
\title{Kaon Multiplicities of Semi-inclusive DIS and the Fragmentation Functions}
\begin{abstract}
There is an apparent discrepancy between the results of the charged kaon multiplicities off the deuteron target from HERMES and COMPASS experiments.
In this article we point out that this discrepancy cannot be explained by different $Q^2$ values.
Furthermore we examine the empirical parametrization of the fragmentation functions, DSS2017 carefully
and find that the agreement between the theoretical estimate and the HERMES data is less satisfactory as claimed.
\end{abstract}

\maketitle
%--------------------------------------------------
\section{Introduction}
Recently there are several attempts to extract the strange-quark Paton Distribution Functions (PDFs)
from the data of the kaon multiplicity of semi-inclusive deep-inelastic scattering (SIDIS) off the deuteron target~\cite{strange1,strange2,strange3},
based on the recent data from HERMES collaboration at DESY ~\cite{Airapetian:2012ki,Airapetian:2013zaw}.
The idea is the strange and anti-strange quarks inside the proton will
hadronize into the charged kaons after being knocked out by the photon with high virtuality.
The leading order (LO) formula of the kaon multiplicity off the deuteron target
is given as,
\begin{eqnarray}
&&M^{K}_{D}(x,Q^2)\equiv M^{K^{+}}_{D}(x,Q^2)+M^{K^{-}}_{D}(x,Q^2)=\frac{dN^{K}(x,Q^2)}{dN^\mathrm{DIS}(x,Q^2)}
\cr
&=&\frac{\sum_{q}e_{q}^2\left[q^{p}(x,Q^2)+\bar{q}^{p}(x,Q^2)+q^{n}(x,Q^2)+\bar{q}^{n}(x,Q^2)\right]\int^{z_\mathrm{max}}_{z_\mathrm{min}}D_{q}^{K}(z,Q^2)dz}
{\sum_{q}e_{q}^2\left[q^{p}(x,Q^2)+\bar{q}^{p}(x,Q^2)+q^{n}(x,Q^2)+\bar{q}^{n}(x,Q^2)\right]}.
\label{Eq:ori2}
\end{eqnarray}
Here $q=(u,d,s)$ and $e_q$ are the quark flavours and the corresponding electric charges, respectively.
$q^{i}(x,Q^2)$ with $i \in \{p,n\}$ are the relevant nucleon PDFs which are the functions of momentum fraction $x$ and
momentum transfer squared $Q^2$. Notice the superscripts $p$ and $n$ denote proton and neutron.
The $z$ is the momentum fraction of the initial quark in the fragmented hadron and $z_\mathrm{max}$ and $z_\mathrm{min}$ are usually set by the
experimental acceptance. Finally $D_{q}^{K}$ is defined as
$D_{q}^{K}(z,Q^2)=D_{q}^{K^{+}}(z,Q^2)+D_{q}^{K^{-}}(z,Q^2)$. From Eq.~(\ref{Eq:ori2}) one would extract
the sum of the strange and the anti-strange quark PDFs provided that the isospin symmetry is assumed.
However, in our previous work we find that such an extraction crucially depends on the choice of the fragmentation functions \cite{Yang:2015avi}.
Furthermore we also point out that such an extraction actually has carried out on the pion multiplicities data, and there is serious tension between
the results from the pion multiplicity and kaon multiplicity~\cite{Yang:2015avi}.
So far all these studies have been based on the leading order (LO) formula, hence it is necessary
to investigate the hadron multiplicities according to the next-leading-order (NLO) formula.
The NLO formula of SIDIS is as follows,
\begin{eqnarray}
\sigma^{h}(x,z)&=&\sum_{f}e_f^2 q_{f}\otimes D^{h}_{q_f}
\cr
&+&\frac{\alpha_s}{2\pi}\left(e_f^2 q_{f}\otimes {\cal C}_{qq}\otimes D^{h}_{q_f}
+e_f^2 q_{f}\otimes {\cal C}_{qg}\otimes D^{h}_{G}
+\frac{\alpha_s}{2\pi}G\otimes {\cal C}_{gq}\otimes \sum_{q_f}e_f^2D^{h}_{q_f}\right).
\cr
&& q\otimes{\cal C}\otimes D(x,z)\equiv \int_{x}^{1}\frac{dx'}{x'}\int_{z}^{1}\frac{dz'}{z'}q\left(\frac{x}{x'}\right){\cal C}(x',z')D\left(\frac{z}{z'}\right).
\label{Eq:NLO}
\end{eqnarray}
Here ${\cal C}$ are the splitting functions given in \cite{Furmanski:1981cw}. In principle, the charged meson multiplicity of SIDIS
provides an excellent source for the flavour separation of the fragmentation functions.
A successful parametrization of the fragmentation functions is expected to explain satisfactorily
the HERMES data of the charged kaon multiplicity through Eq.~(\ref{Eq:NLO}). However, one meets some unexpected predicament
when attempting to achieve this goal, we will explain this dilemma in the next section.\\

\section{The Kaon multiplicities of HERMES and COMPASS Data}
\begin{figure}
\begin{tabular}{cc}
\includegraphics[width=.5\textwidth]{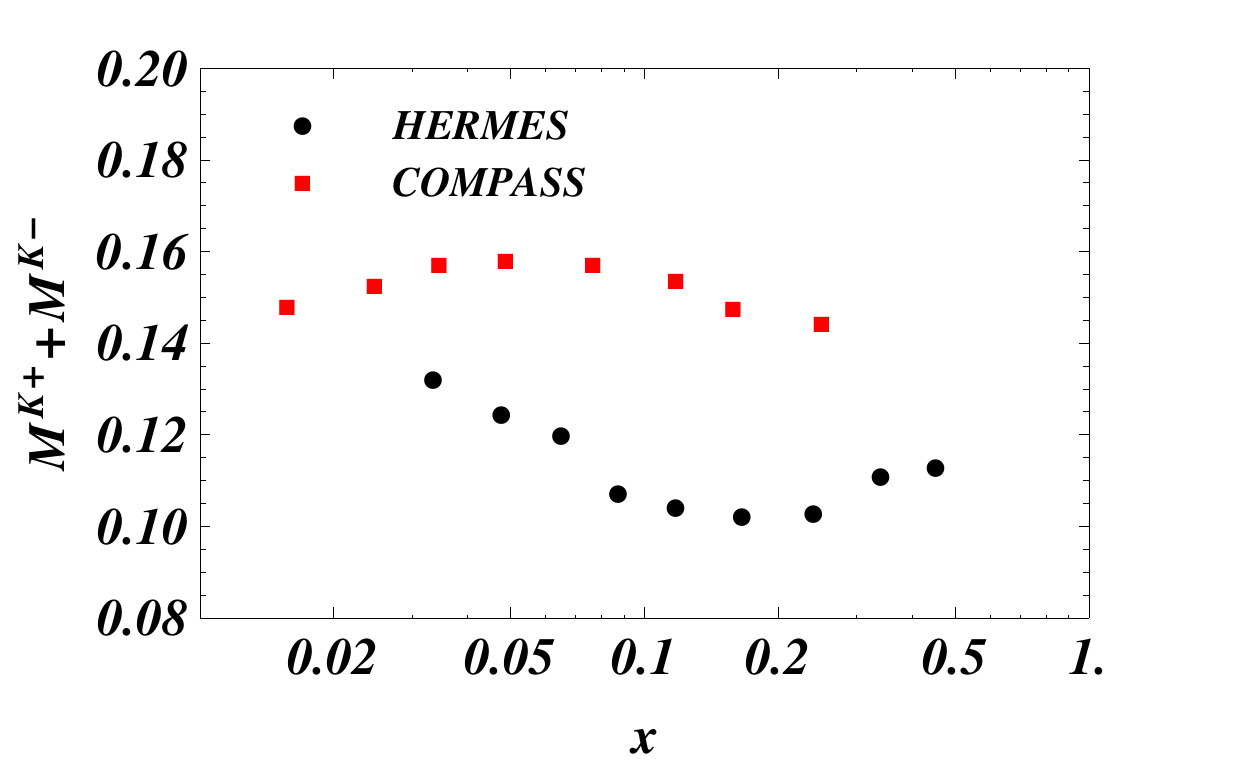}
\includegraphics[width=.5\textwidth]{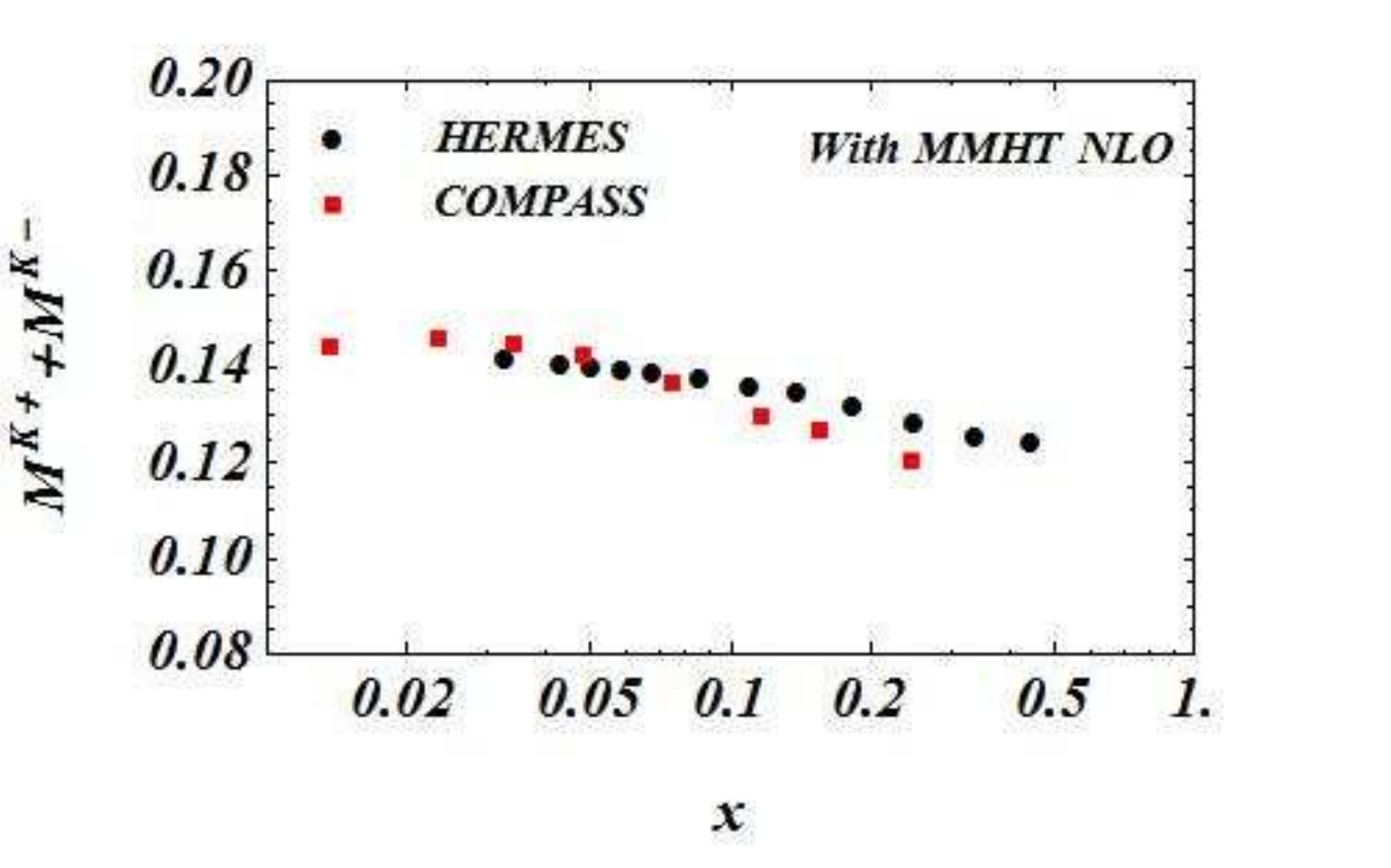}
\end{tabular}
\caption{(a)Experimental results of charged kaon multiplicities from HERMES ~\cite{Airapetian:2012ki,Airapetian:2013zaw} and COMPASS ~\cite{Adolph:2016bwc} experiments.
(left panel). (b) Theoretical predictions with the DSS2017 parametrization and MMHT PDFs at HERMES kinematics compared with the experimental data.(right panel)}
\label{data}
\end{figure}

Recently the COMPASS results of the charged
kaon multiplicities of SIDIS off the deuteron target have been published ~\cite{Adolph:2016bwc}.
From Fig.(\ref{data}-a) the COMPASS results obviously deviate from the HERMES ones. However such a difference may be
just superficial because the corresponding $Q^2$ of two sets of data are different.
To assert the inconsistency between two data sets, one needs to exclude the possibility of that
this difference is merely caused by the  QCD evolution between different $Q^2$ values.
However it is difficult, if not entirely impossible, to apply QCD evolution
on the experimental data directly. Instead, it is more practical to choose reliable FFs and PDFs and apply the NLO formula
, Eq.~(\ref{Eq:NLO}) to obtain the theoretical predictions at the precise $Q^2$ scales.\\

 In 2017 DSS2017 Kaon fragmentation functions became available \cite{deFlorian:2017lwf}. This parametrization is the updated version of the previous one~\cite{deFlorian:2007ekg}. The major updating is to include the COMPASS data~\cite{Adolph:2016bwc}. Their differential kaon multiplicity predictions are compared with the COMPASS and HERMES data. It is claimed their parametrization is able to describe both of the data sets well simultaneously ~\cite{deFlorian:2007ekg}. Therefore we use DSS2017 parametrization with MMHT PDFs~\cite{Harland-Lang:2014zoa} (as used in~\cite{deFlorian:2007ekg}) and present our results in Fig.~(\ref{data}-a). We demonstrate the NLO result at the HERMES and COMPASS kinematics, respectively. It is obvious that their agreement with the experimental data is far from satisfactory. Furthermore, the discrepancy between the two data is not from the $Q^2$
difference since the theoretical results of both are very close as shown in Fig.~(\ref{data}-b).
For further investigation, one needs to clarify the issue that whether DSS2017 parametrization can describe the COMPASS as well as HERMES data
in term at differential charged kaon multiplicities~\cite{deFlorian:2007ekg}.
Since DSS2017 paper demonstrates their result agree with the experimental data well before integrating $z$ values,
therefore, it is necessary to examine the DSS2017 parametrization more intensively.

\section{The differential charged kaon multiplicities from the DSS2017 parametrization}

\begin{figure}
\begin{tabular}{cc}
\includegraphics[width=.4\textwidth]{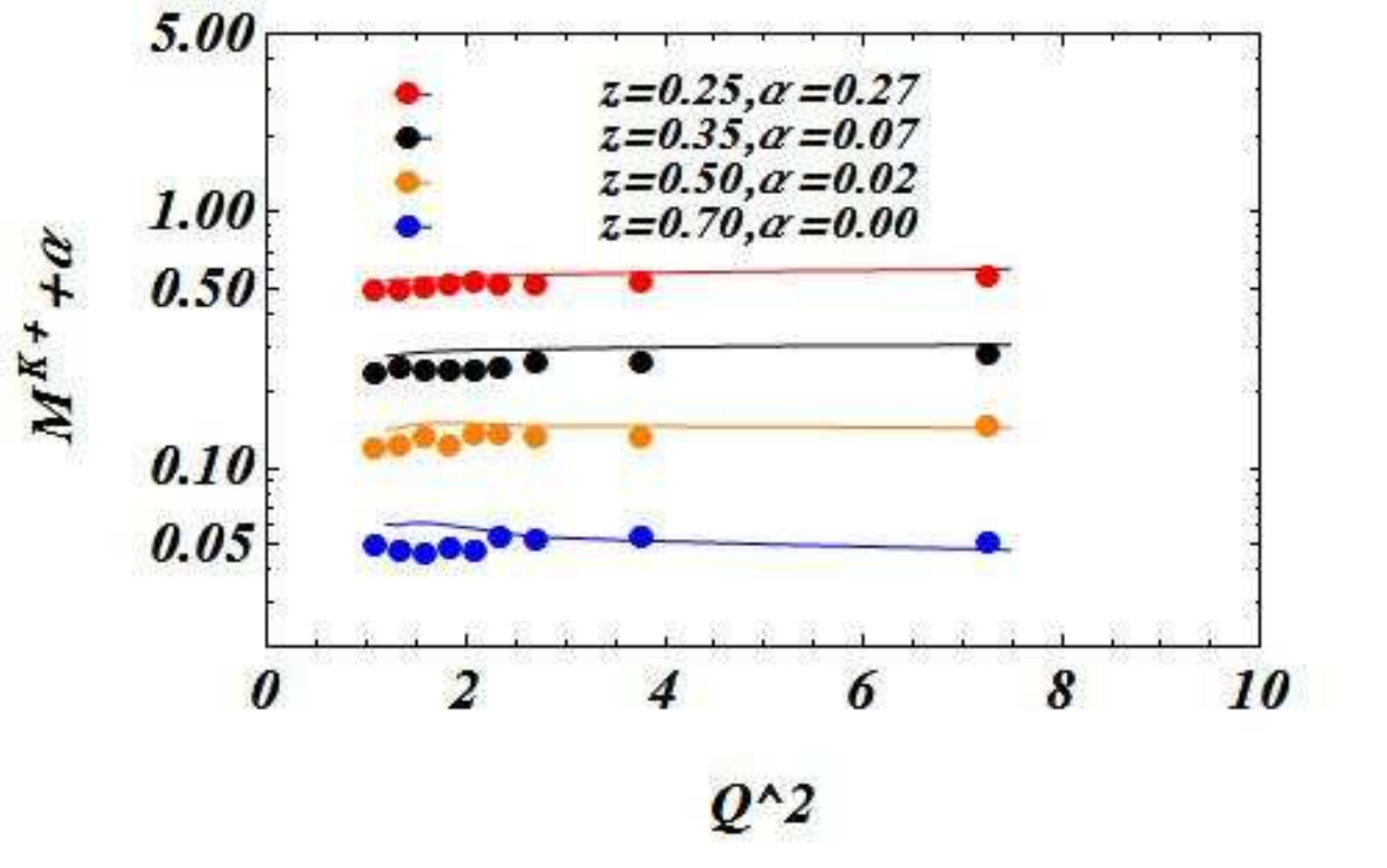}
\includegraphics[width=.4\textwidth]{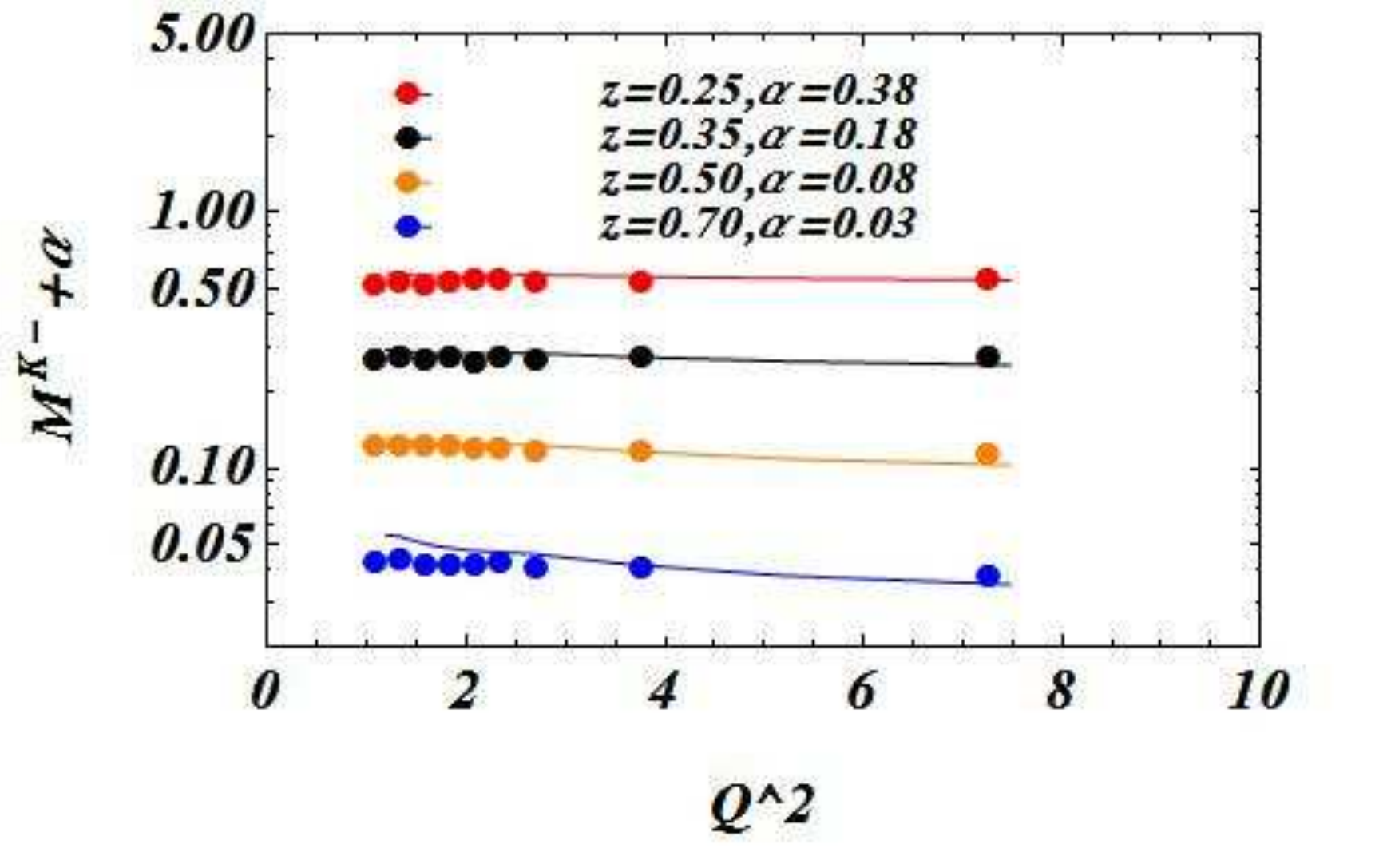}
\end{tabular}
\begin{tabular}{cc}
\includegraphics[width=.4\textwidth]{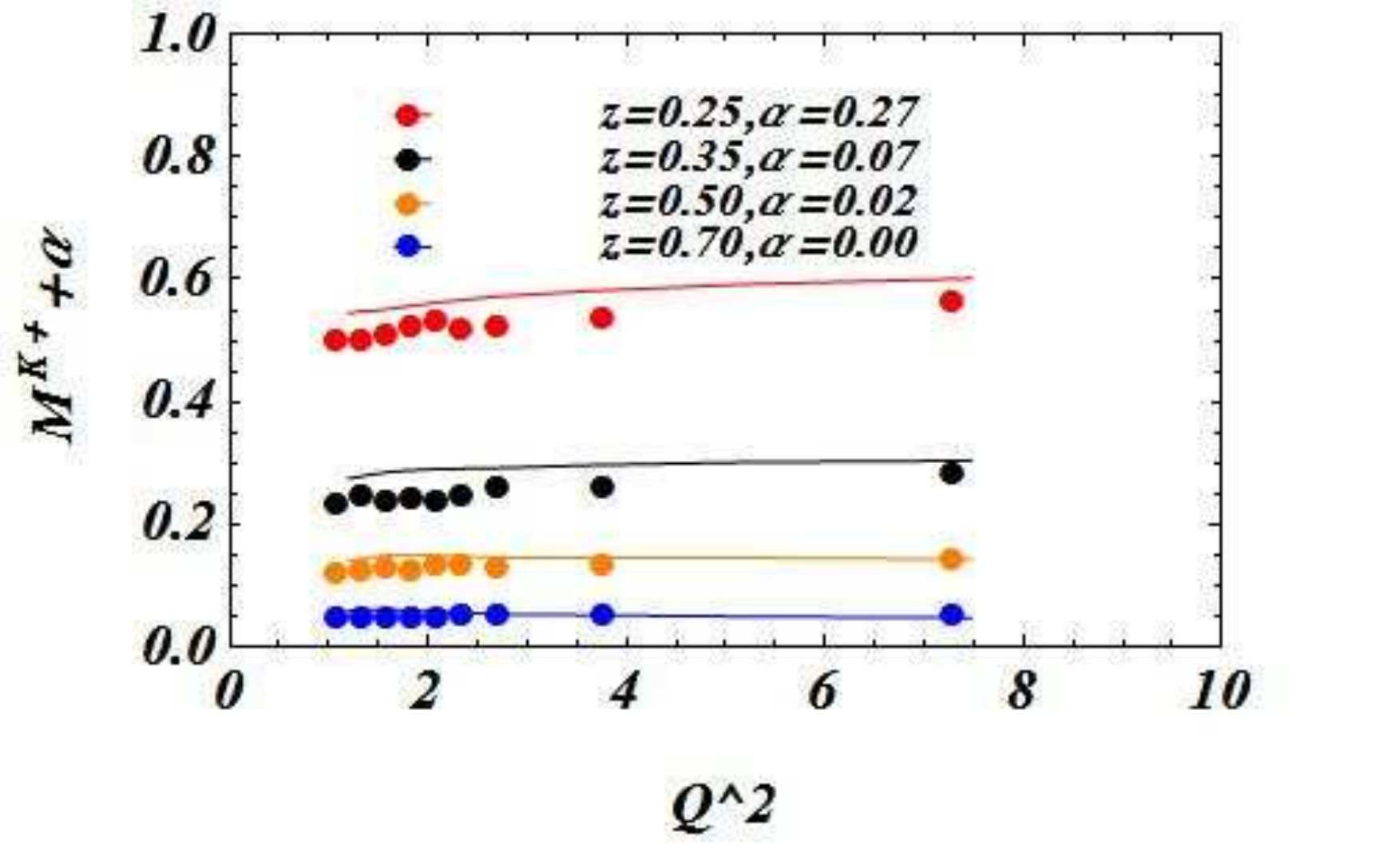}
\includegraphics[width=.4\textwidth]{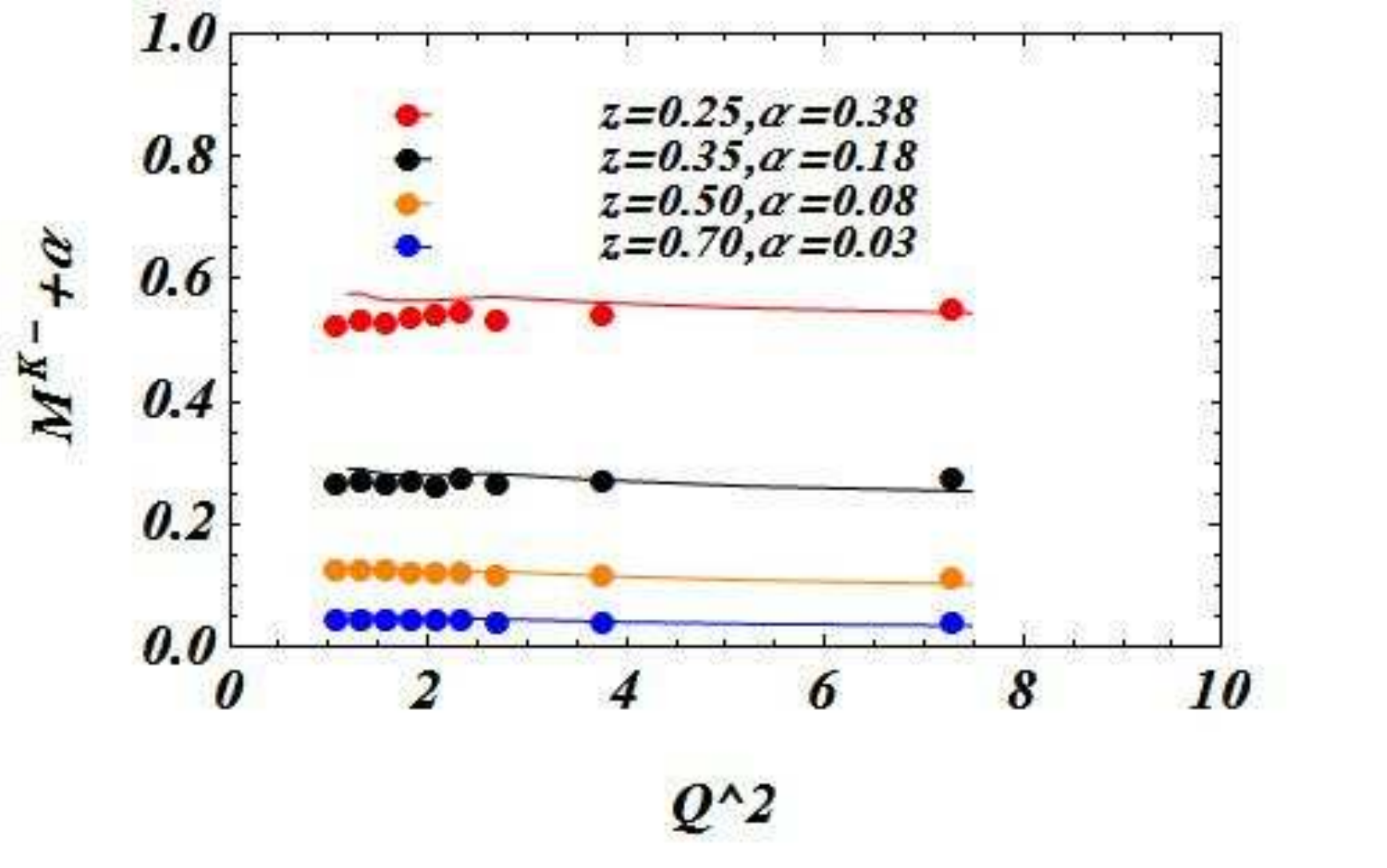}
\end{tabular}
\caption{The kaon multiplicities at different $z$ ranges. The $y$ axis is logarithm scale for the first row of figures.
For the second row of the figures, the $y$ scale is linear.}
\label{log}
\end{figure}

Using DSS2017 \cite{deFlorian:2017lwf} we obtained the kaon multiplicities off the deuteron shown in Fig.(\ref{log}). The $\alpha$ is just added
to shift the results of the different $z$ values vertically to make the comparison more transparent.

In Fig~(\ref{log}), the top two plots have shown that the agreement between the DSS2017 differential result and the HERMES experimental data seems excellent.
Unfortunately it is mere illusory. When one changes the logarithmic scale into the linear one, the seemingly excellent agreement is gone.
One can easily figure it out by observing the bottom two of plots in Fig.~(\ref{log}).
The curves in ~Fig.~(\ref{log}) are just the lines connecting the theoretical result at each data point. It turns out that the vertical shift the data creates a visionary effect which is rather misleading since the size of the data point does not reflect the experimental uncertainty properly. The data points shifted to the higher $y$ values own a much larger error range than the data points which are not shifted since the $y$ axis is logarithmic, however,
the authors in~\cite{deFlorian:2017lwf} did not adjust the size of the data points according to their experimental uncertainty.\\

\begin{figure}
\begin{tabular}{ccc}
\includegraphics[width=.3\textwidth]{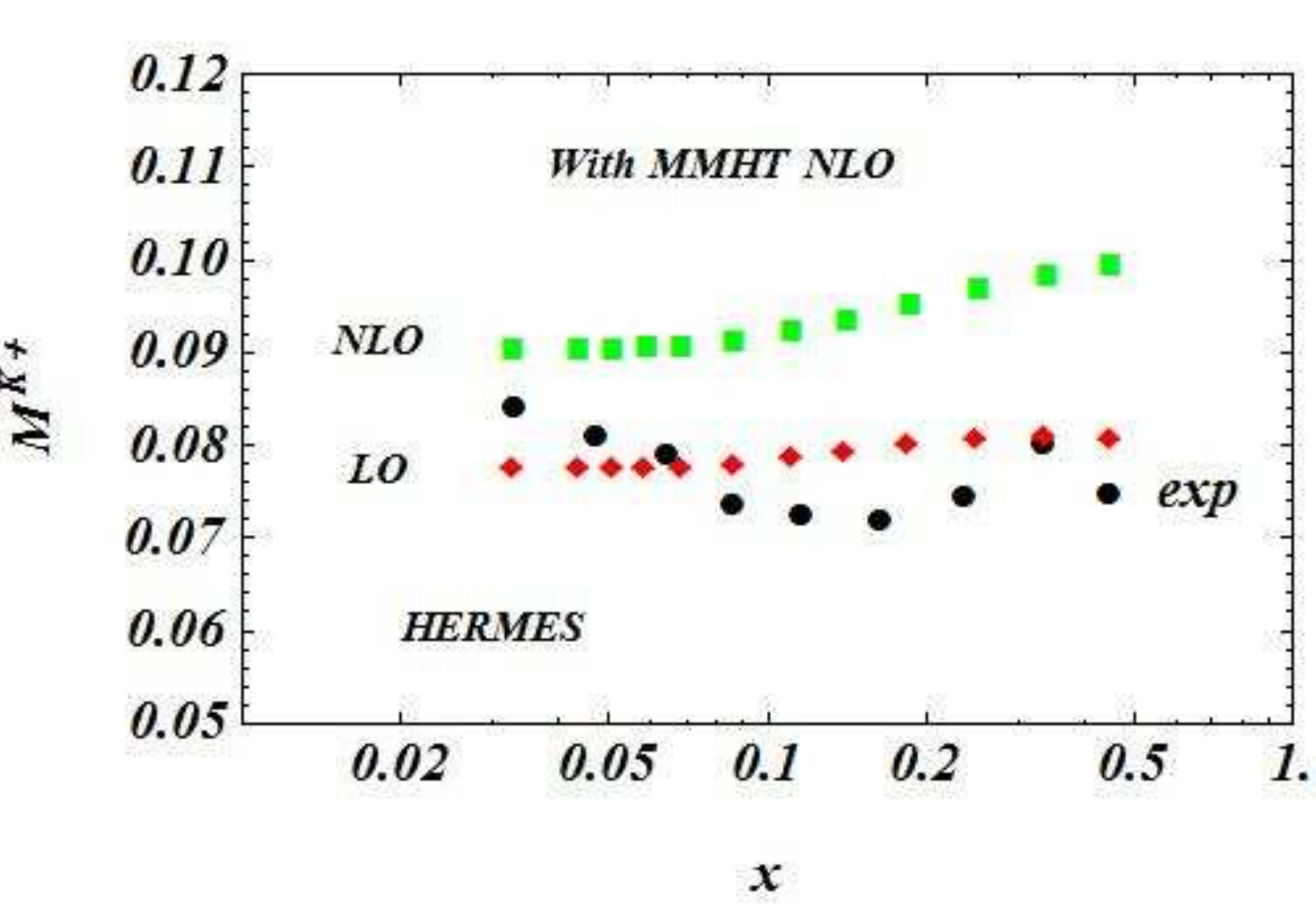}
\includegraphics[width=.3\textwidth]{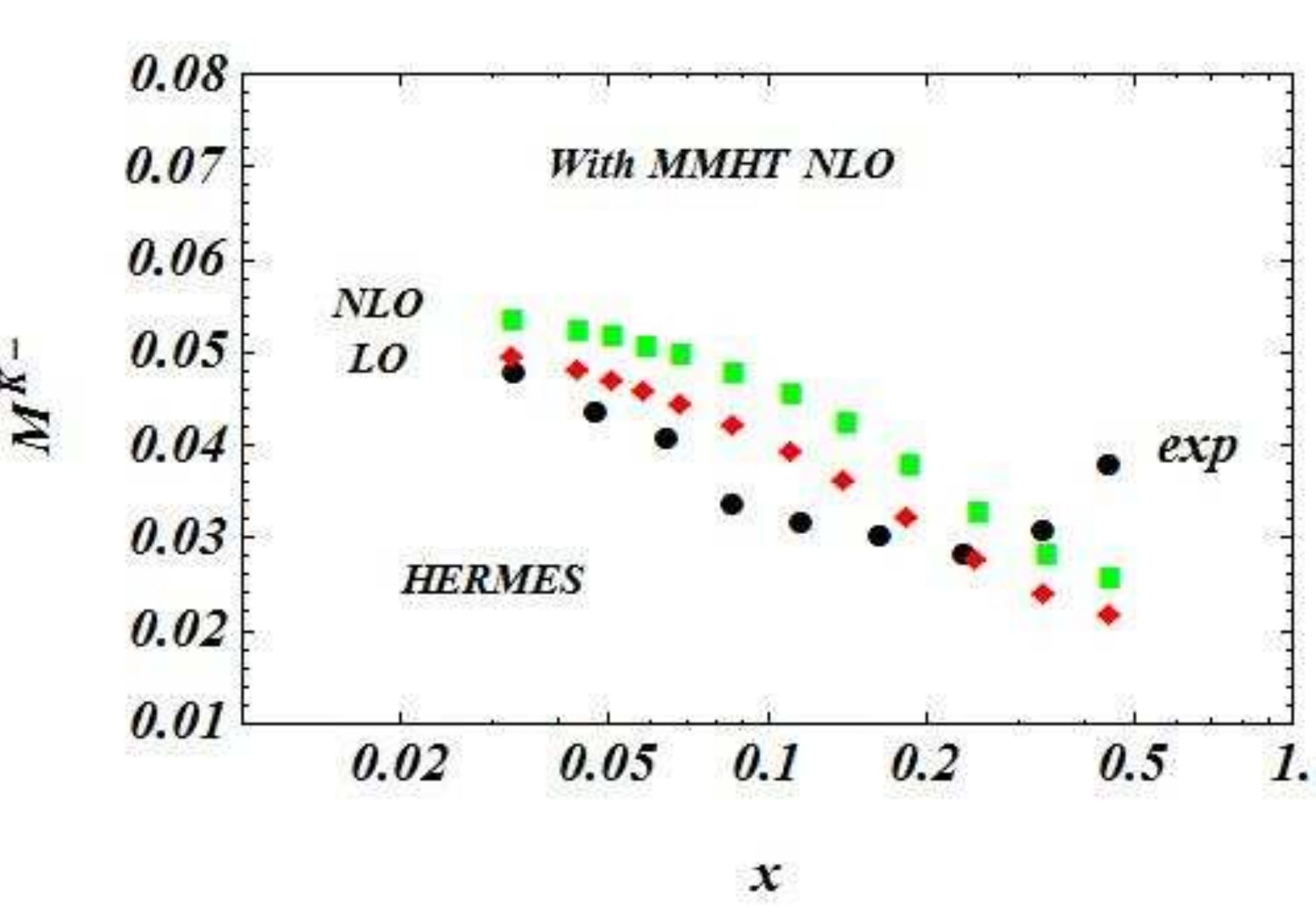}
\includegraphics[width=.3\textwidth]{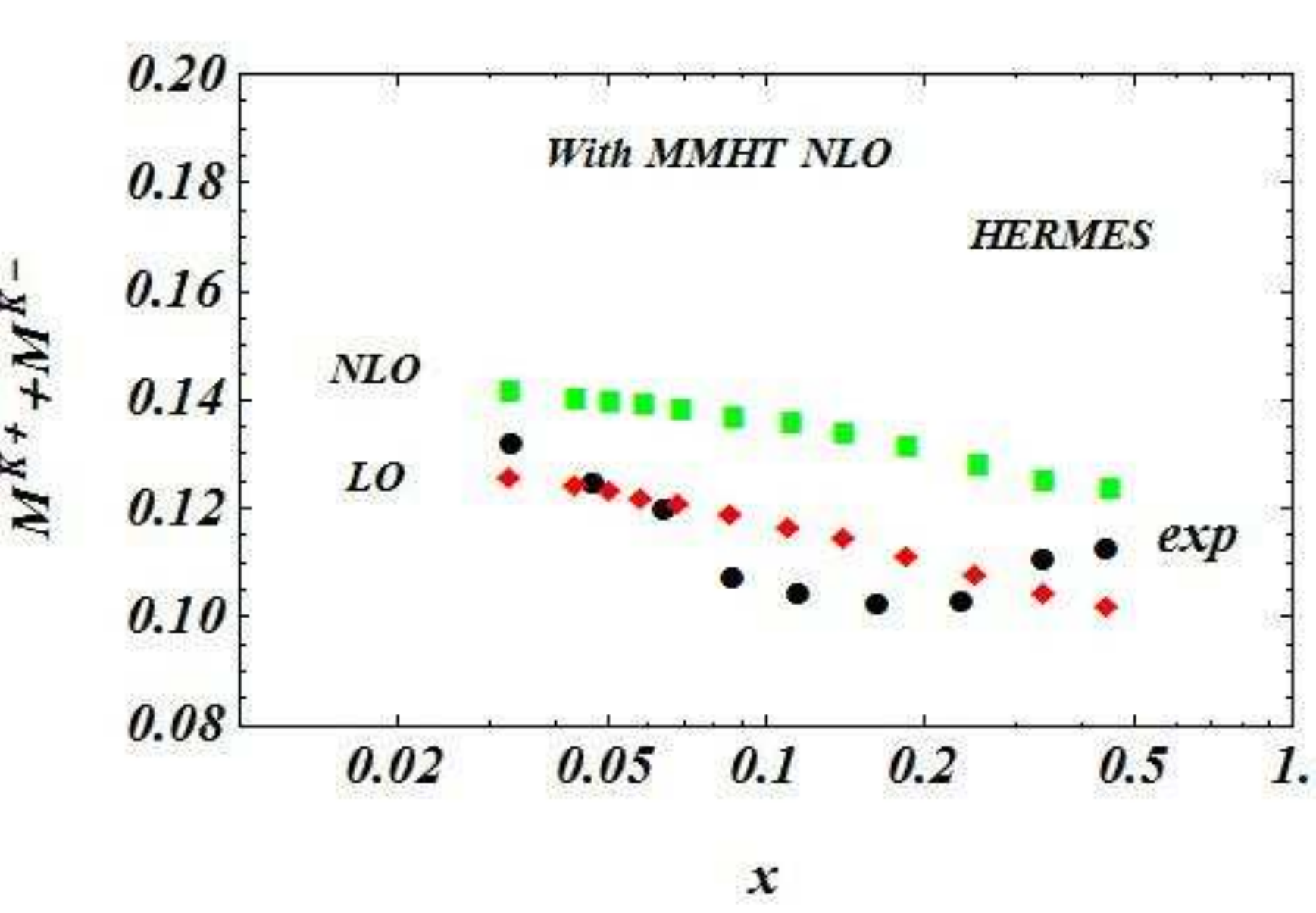}
\end{tabular}
\label{sum}
\caption{Comparison between the theoretical and experimental results of the kaon multiplicities at the kinematics of HERMES data.
(a) $K^{+}$ multiplicity (b) $K^{-}$ multiplicity (c) The sum of $K^{+}$ and $K^{-}$ multiplicity.  }
\end{figure}
From Fig.~(3-a) one find the theoretical curves are larger than the experimental data about $10\%$ when $z$ is smaller than $0.35$
in the $K^{+}$ case. This trend is valid for all the data points even they are taken at different $x$ and $Q^2$ values.
In the $K^{-}$ case, the differences between the curve and the data become smaller and at the large $z$ regime the data is even larger than the curve
as shown in Fig.~(3-b). Summing over $z$ range one obtains the results presented in Fig.~(3-c).
It turns out that the difference between the theoretical result and the HERMES data is most significant in the middle-$x$ region with $Q^2$ located between
$2-4$ GeV$^2$. The difference would reach $40\%$!

\section{Conclusion}
In general the HERMES results of the kaon multiplicity are smaller than what one expects
if DSS2017 parametrization is used. Hence the claim that the DSS2017 parametrization is able to describe both of the data sets well simultaneously in~\cite{deFlorian:2007ekg}
is not confirmed in our study. Besides we find that the NLO contribution of the charged kaon multiplicity actually makes the agreement between the theoretical estimate and the experimental data worse as shown in Fig.~(3).
%\begin{figure}
%\includegraphics[width=.6\textwidth]{Fig6.eps}
%\caption{The comparison between the theoretical and experimental results of the charged kaon multiplicities at the kinematics of HERMES data. }
%\label{total}
%\end{figure}

%\section{Conclusion and Outlook}
% We investigate the charged kaon multiplicities off the deuteron target from HERMES and COMPASS experiments. The discrepancy of the two data
% cannot be explained by the different $Q^2$ values.
% Furthermore we find that the agreement between the NLO theoretical predictions with the DSS2017 parametrization
% and the HERMES data are less satisfactory as claimed.
% We plan to study the similar issues for the COMPASS data and extend our analysis to other hardon multiplicities of the SIDIS off the deuteron and proton target and hope our % % study will shed some light on the cause of the discrepancy of the HERMES and COMPASS data.

\end{document}